\documentclass[11pt,a4paper]{scrartcl}

\usepackage{ILD}

\usepackage[bottom]{footmisc}
\usepackage{feynmf}
\usepackage{multirow}
\usepackage{textpos}

\usepackage{hyperref}
\usepackage{url}
\usepackage{mathtools}
\usepackage{wrapfig}
\usepackage{footnote}
\usepackage{subcaption}
\usepackage{caption}
\usepackage{verbatim}
\usepackage{comment}
\usepackage{indentfirst}
\usepackage{wasysym}
\usepackage{multirow}
\usepackage{xspace}
\usepackage{acronym}

\newcommand{\epem}{e$^+$e$^-$\xspace}

\newlength{\figwidth}
\newlength{\figskip}
\title{Reconstructing long-lived particles with the ILD detector}

\ildproc{phys}{2023}{010} 

\date{\today}

\addauthor{Jan Klamka}

\addinstitute{}{Faculty of Physics, University of Warsaw,\\
  Pasteura 5, 02-093 Warsaw, Poland}

\abstract{
Future \epem colliders, thanks to their clean environment and triggerless operation, offer a unique opportunity to search for long-lived particles (LLPs). Considered in this contribution are promising prospects for LLP searches offered by the International Large Detector (ILD), with a Time Projection Chamber (TPC) as the core of its tracking systems, providing almost continuous tracking. The ILD has been developed as a detector concept for the ILC, however, studies on understanding of the ILD performance at other collider concepts are ongoing.

Based on the full detector simulation, we study the possibility of reconstructing decays of both light and heavy LLPs at the ILD. For the heavy, $\mathcal{O}$(100 GeV) LLPs, we consider a challenging scenario with small mass splitting between the LLP and the dark matter candidate, resulting in only a very soft displaced track pair in the final state, not pointing to the interaction point. We account for the soft beam-induced background (from measurable \epem pairs and hadron photo-production processes), expected to give the dominant background contribution due to a very high cross section, and show the possible means of its reduction. As the opposite extreme scenario we consider the production of a light, $\mathcal{O}$(1 GeV) pseudo-scalar LLP, which decays to two highly boosted and almost colinear displaced tracks.
We also present the corresponding results for an alternative ILD design, where the TPC is replaced by a silicon tracker modified from the Compact Linear Collider detector (CLICdet) design.
}

\graphicspath{ {./logos/}{./figures/} }

\begin{document}

\titlepage

\section{Introduction}

The concept of Beyond the Standard Model (BSM) particles with macroscopic lifetimes has been extensively studied in recent years, also in the context of collider searches. The properties needed for a particle to acquire a significant lifetime are either a small coupling, compressed spectrum of mass states, or a scale suppression. Hence, LLPs can naturally appear in a variety of the BSM scenarios~\cite{Lee:2018pag}. These properties also suggest, that an \epem Higgs factory could be a good place for the LLP searches due to its clean experimental environment and trigger-less operation. 

The International Large Detector (ILD)~\cite{ILDConceptGroup:2020sfq} is a multipurpose detector concept for an experiment at the Higgs factory, originally proposed for the International Linear Collider (ILC)~\cite{Bambade:2019fyw}. Its design is optimised for the particle-flow reconstruction and provides almost $4\pi$ angular coverage. The core of the ILD tracking systems is a time projection chamber (TPC) which allows for almost continuous tracking and dE/dx measurement for particle identification. As most of LLP signatures can be categorised as tracks that do not point to, or do not originate from the interaction point (IP), ILD with its TPC provides very interesting prospects for these measurements. 

An experiment-focused approach is taken here. Benchmark scenarios are selected based on their experimental/kinematic properties, and not as preferred points in a model parameter space. The signature of two tracks coming from a displaced vertex is studied as a generic case of an LLP search. For generality, no other assumptions about the final state are made.

\section{Analysis framework}

Two opposite extreme scenarios are considered as the benchmarks. The more challenging case involves the production of heavy scalars with small-boost and non-pointing, low-$p_T$ track pair in the final state. This is provided by the Inert Doublet Model (IDM)~\cite{Kalinowski:2018ylg}, which introduces four additional scalars, including two neutral ones (one of them, H, being a stable dark matter candidate). The neutral scalars, A and H, can be produced in \epem $\to$ AH process, with A $\to$ Z$^{(*)}$H as the main decay channel. If the mass splitting, $\Delta m_{AH}=m_A - m_H$, between the two scalars is sufficiently small, A could be long-lived. This also makes the Z boson highly virtual, so for high masses of the scalars (small boost), the observed decay products will be soft and will not point to the IP. Four IDM benchmarks are selected for the analysis, with $\Delta m_{AH}=1,2,3,5$\,GeV, decay length $c\tau$ of A fixed to 1\,m, and its mass $m_A = 155$\,GeV. Only Z$^{*}$ decays to muons are considered.

The second type of scenario under study is a production of a very light pseudoscalar, the decay products of which have a very high boost and are almost colinear. This is the case for an axion-like particle (ALP), which in many scenarios has a macroscopic lifetime for masses of the order of GeV, and could also be produced at \epem colliders in \epem $\to$ a$\gamma$ channel~\cite{Schafer:2022shi}. The four analysed benchmarks assume different ALP masses, $m_a = 0.3, 1, 3$ and $10$\,GeV, and decay lengths $c\tau = 10\cdot m_a$\,mm/GeV to maintain a consistent number of decays within the detector volume. Again, only ALP decays to muon pairs are considered.


\textsc{Whizard} 3.0.1 and 3.1.2~\cite{Kilian:2007gr} was used to generate IDM at 500\,GeV and ALP samples at 250\,GeV ILC, respectively, while the detector response simulation was based on \textsc{Geant4}~\cite{AGOSTINELLI2003250}. For the vertex finding, an algorithm which is both simple and as general as possible has been developed. It is based on the calculated distance between the track helices (which is required to be 25\,mm at most) and the vertex is placed in between the points of the closest approach of the helices. At this stage a basic preselection is applied, aimed to reduce the number of fake vertices in the signal sample. The cuts are applied based on track curvature, length, opening angle, and distance from the first hit to vertex.

\section{Background and its reduction}

At linear colliders, beam-induced background events can contribute in each bunch-crossing (BX). On average, 1.05 events of the $\gamma\gamma\to$ hadrons process and a fraction of produced incoherent \epem pairs can be measured in the detector per single BX at the ILC. In most physics studies, this background can \textit{overlay} on a hard event and affect reconstruction results. However, when low-$p_T$ final states are expected in the signal channel considered, these events can constitute background on their own and must be considered separately.

\begin{figure}[bt]
	    \centering
	 	 \begin{subfigure}{0.49\textwidth}
	 	 	\centering
	 	 	\includegraphics[width=\figwidth]{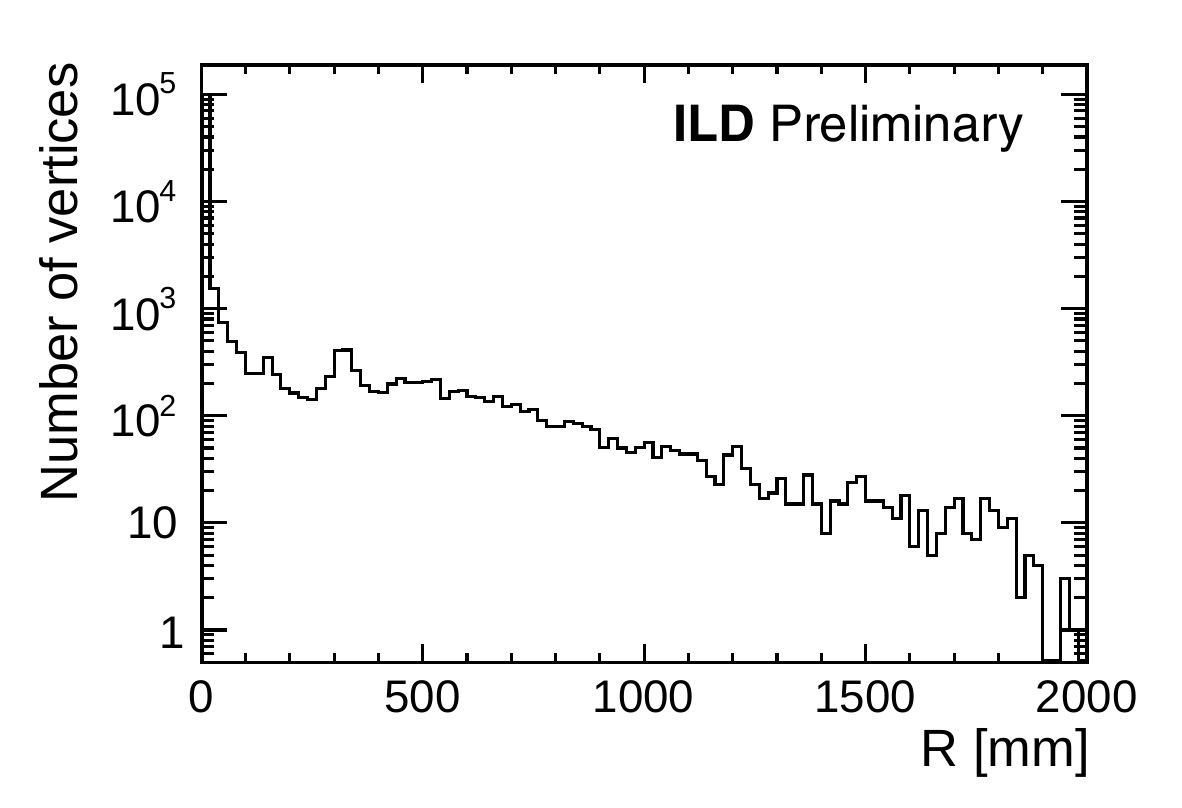}
	 	 \end{subfigure}%
	 	 \begin{subfigure}{0.49\textwidth}
	 	 	\centering
	 	 	\includegraphics[width=\figwidth]{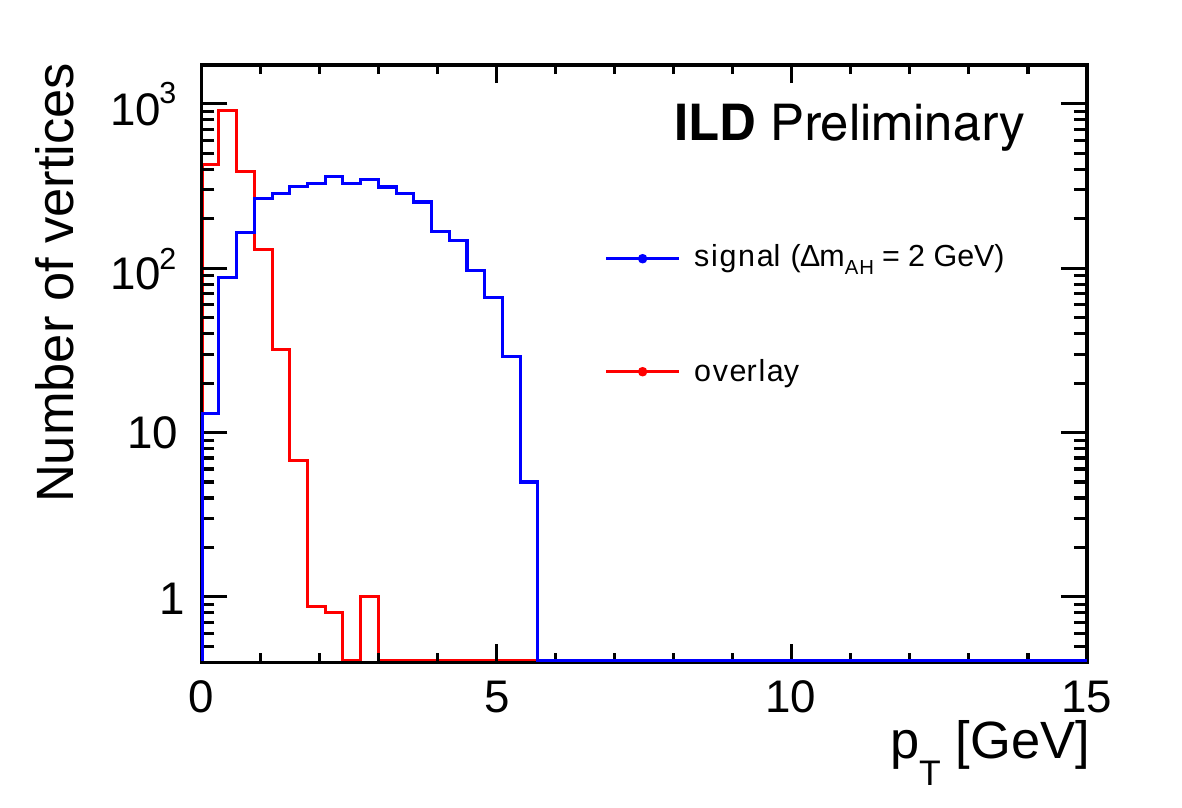}
	 	 \end{subfigure}
  \vspace*{\figskip}
	 	 \caption{Left: Number of displaced vertices found in the overlay sample as a function of distance from the beam axis with no selection applied at all, just apart from the required maximum distance between track helices. Right: Total transverse momentum of tracks coming from a displaced vertex for the overlay (red) and one of the signal scenarios (blue; heavy scalars production with $\Delta m_{AH}=2$\,GeV) after imposing preliminary selection cuts.  }
	 	 \label{fig:background}
\end{figure}

This is also the case for the presented analysis, with the minimal number of assumptions about the final state and the universal approach to the vertex finding. Overlay events can contain a sizable number of very low-$p_T$ tracks that start near the IP. In such a busy environment of curling tracks, the algorithm designed for the purpose of this study finds many (mostly fake) vertices, especially close to the beam axis. This is visible in Fig.~\ref{fig:background} (left), where the number of vertices is presented as a function of distance from the beam axis (without any preselection cuts imposed). Tracks at the vertices in overlay events can also be kinematically similar to those in the analysed signal scenarios. In Fig.~\ref{fig:background} (right) $p_T$ of the track pair coming from a displaced vertex is presented, for the heavy scalar production scenario (with $\Delta m_{AH}=2$\,GeV) and for the overlay background, after preselection. One can see that both signal and background occupy the same kinematic region. Considering that the number of BXs expected at the ILC is of the order of $10^{11}$, the overlay becomes a very significant background.

However, it is possible to strongly suppress the overlay background even in the presented model-independent approach. The selection involves cuts on the pair transverse momentum, $p_T>1.9$\,GeV, on distances between first hits in tracks and between centres of ``helix-circles" (track projections into XY plane). Together with preselection, this provides a total rejection factor at the level of $10^{9}$ ($10^{10}$) for $\gamma\gamma\to$ had. (\epem pairs).

\section{Results in the TPC}

\begin{figure}[bt]
	    \centering
	 	 \begin{subfigure}{0.49\textwidth}
	 	 	\centering
	 	 	\includegraphics[width=\figwidth]{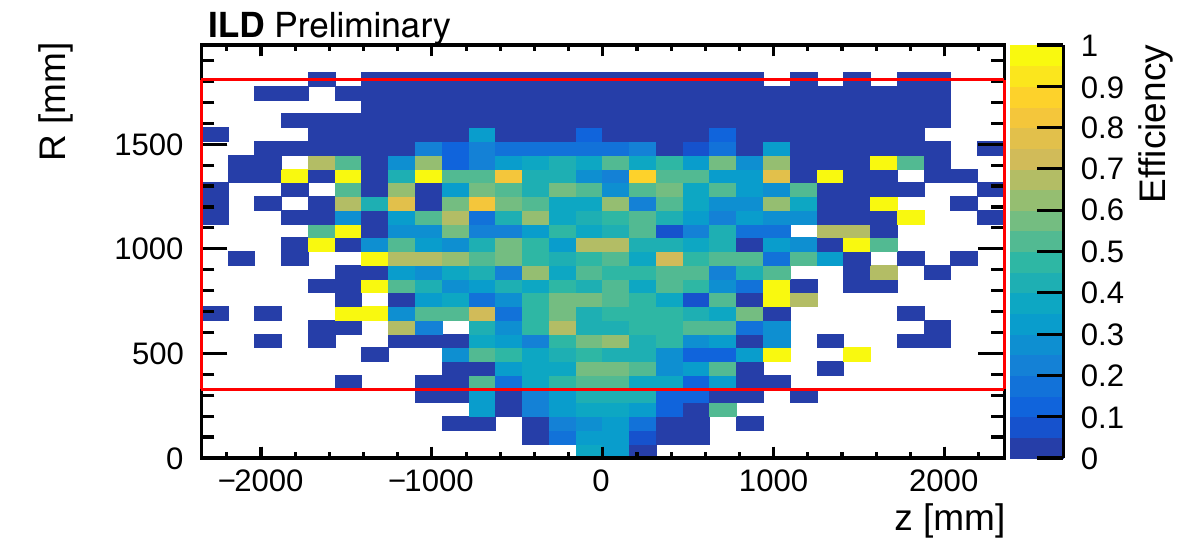}
	 	 \end{subfigure}%
	 	 \begin{subfigure}{0.49\textwidth}
	 	 	\centering
	 	 	\includegraphics[width=\figwidth]{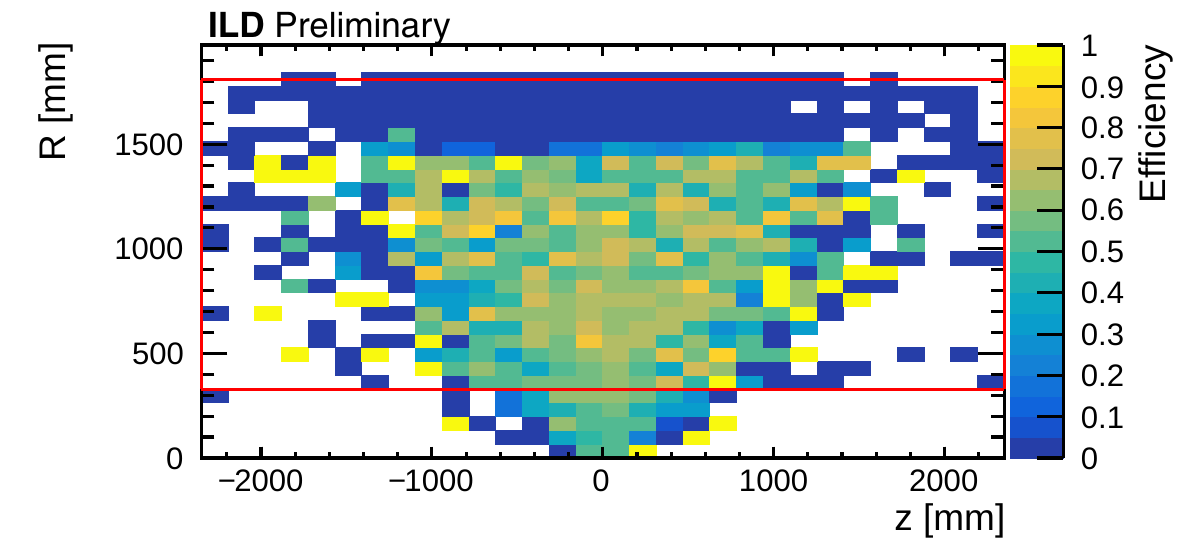}
	 	 \end{subfigure}
	  \vspace*{\figskip}
 	 \caption{Vertex finding efficiency as a function of the true LLP decay vertex position in the detector, for two mass splittings, $\Delta m_{AH}=2$\,GeV (left) and $\Delta m_{AH}=3$\,GeV (right), in the heavy scalars (IDM) scenario. The TPC volume is shown with the red box.}
	 	 \label{fig:res_idm}
\end{figure}

Fig.~\ref{fig:res_idm} shows the resulting vertex finding efficiency for signal events, for two of the heavy scalar scenarios, with  $\Delta m_{AH}=2$\,GeV (left) and $\Delta m_{AH}=3$\,GeV (right), as a function of the true decay vertex position in the detector. The reconstructed vertex was considered to be ``correct'', if its distance from the true vertex was less than 30 mm. The efficiency is also shown for LLP decays before the TPC, and it should be noted that vertex finding efficiency increases when the tracks start inside the TPC volume. 

\begin{figure}[bt]
	    \centering
	 	 \begin{subfigure}{0.49\textwidth}
	 	 	\centering
	 	 	\includegraphics[width=\figwidth]{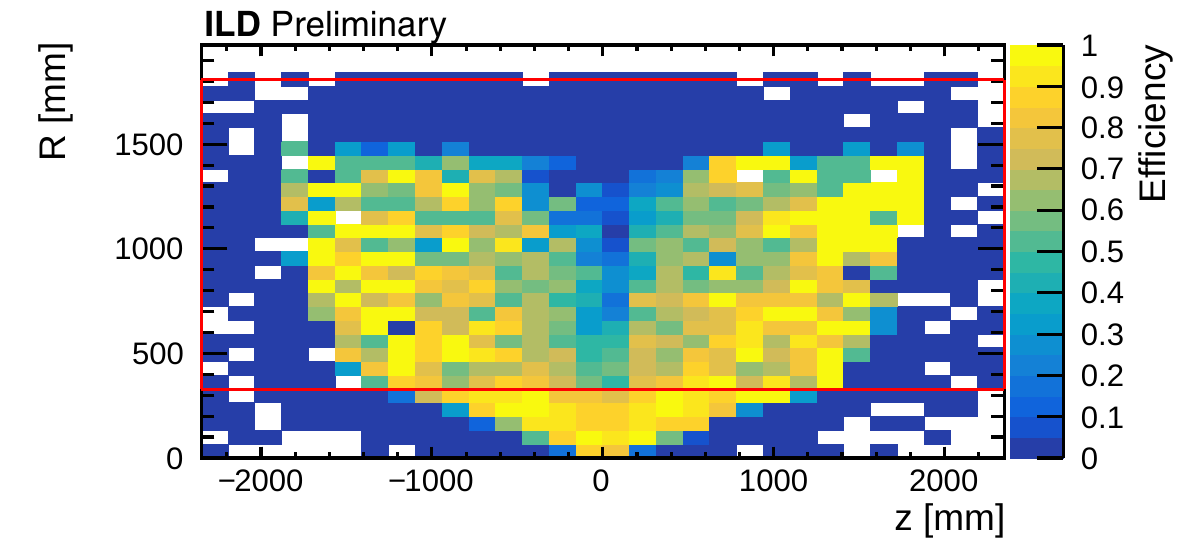}
	 	 \end{subfigure}%
	 	 \begin{subfigure}{0.49\textwidth}
	 	 	\centering
	 	 	\includegraphics[width=\figwidth]{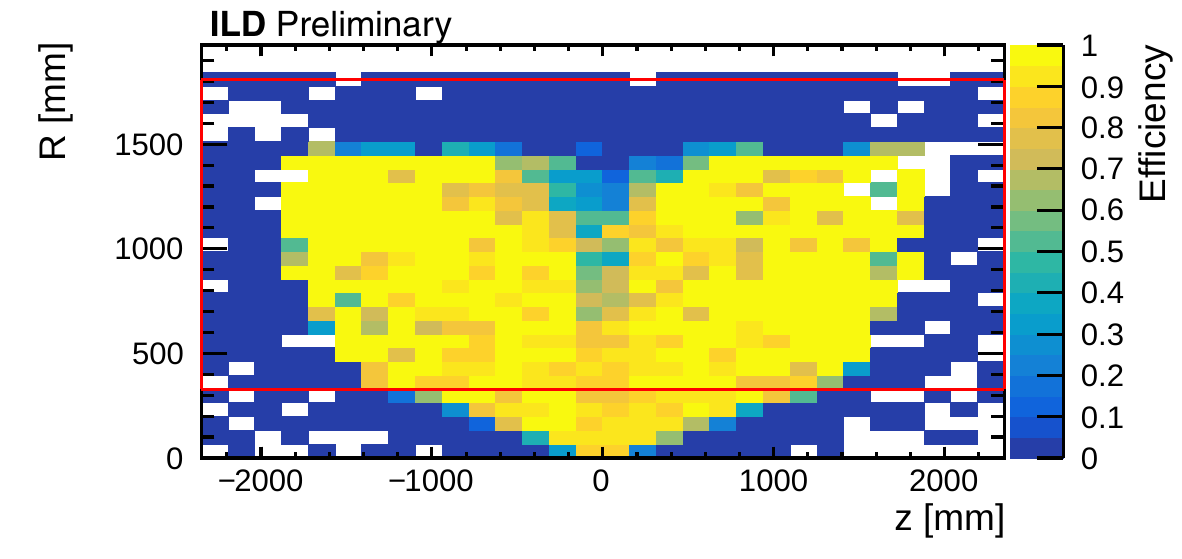}
	 	 \end{subfigure}
	  \vspace*{\figskip}
 	 \caption{Vertex finding efficiency as a function of the true LLP decay vertex position in the detector, for two masses, $m_a = 1$\,GeV (left) and $m_a = 3$\,GeV (right), in the light pseudoscalars (ALPs) scenario. The TPC volume is shown with the red box.}
	 	 \label{fig:res_alp}
\end{figure}

The corresponding efficiency for the light pseudoscalar case is presented in Fig.~\ref{fig:res_alp}, for two masses of ALPs, $m_a = 1$\,GeV (left) and $m_a = 3$\,GeV (right). Contrary to the heavy scalar case, the efficiency is higher in the region of silicon tracker, closer to the beam axis. This is because for high boosts, two tracks in the final state become almost colinear and the separation of hits close to the vertex is better in the silicon detector, providing much higher resolution.

The results on the efficiency and purity of the LLP vertex finding are shown in Table~\ref{tab:results}. Efficiency is defined as the ratio of the number of correctly reconstructed vertices to the number of decays inside TPC, whereas purity is defined as the ratio of the number of correct to all found vertices. The results presented confirm that efficiency strongly depends on the final-state boost. For heavy scalars, good efficiency is achieved already for $\Delta m_{AH}=2$\,GeV and it increases with scalar mass splitting (Z$^\ast$ virtuality), while the purity is almost 100\% for all scenarios. A dedicated approach could also significantly improve results for smaller mass splittings~\cite{Berggren:2013vfa}, such as the $\Delta m_{AH}=1$\,GeV scenario. In the case of light psudoscalar production, high efficiencies and purities are obtained already for  $m_{a}=1$\,GeV and they are increasing with ALP mass (decreasing final state boost and track colinearity). 

\begin{table}[bt]
    \centering
    \caption{The efficiencies and purities obtained in the analysis, both for heavy scalars and light pseudoscalars for all considered scenarios. See text for details.}
    \label{tab:results}
    \begin{tabular}{ccccc} \hline
          $\Delta m_{AH}$ [GeV]&  1&  2&  3& 5\\ \hline
         Efficiency [\%]&  4.2&  38.2&  52.3& 62.1\\ 
         Purity [\%]&  99.1&  99.5&  99.5& 99.7\\ \hline 
         $m_{a}$ [GeV]&  0.3&  1&  3& 10\\ \hline
         Efficiency [\%]&  23.9&  53.8&  76.6& 78\\ 
         Purity [\%]&  42.7&  82.9&  97.4& 99\\ \hline
    \end{tabular} 
\end{table}

With the expected background levels, a rough estimate can be made of the expected 95\% C.L. limit on the signal cross section. Assuming an integrated luminosity of 2\,ab$^{-1}$ at 250\,GeV ILC, corresponding to 10 years of data collection with 10$^{11}$ BXs each year, 1.05 (1.00) $\gamma\gamma\to$ had. (seeable \epem pair) events expected per BX, and the total background rejection of 10$^{-9}$ (10$^{-10}$), one can expect around 1150 background events in total. In the considered kinematic region and TPC volume, this corresponds to the limit on signal cross section $\sigma_{95\% \,\mathrm{C.L.}} \approx 0.03$\,fb (and 0.01\,fb for 4\,ab$^{-1}$ collected at 500\,GeV in 8.5 years).

\section{All-silicon ILD design}

An alternative ILD design was also considered, with the TPC  replaced by the outer silicon tracker from the detector model proposed for the Compact Linear Collider (CLICdet)~\cite{AlipourTehrani:2254048}. A barrel layer had to be added, and the spacing between endcap layers increased, with respect to the CLICdet design to fit the ILD geometry. As for CLICdet, the Conformal Tracking~\cite{Brondolin:2019awm} algorithm was also used as a pattern recognition tool.

\begin{figure}[htb]
    \centering
    \includegraphics[width=0.6\textwidth]{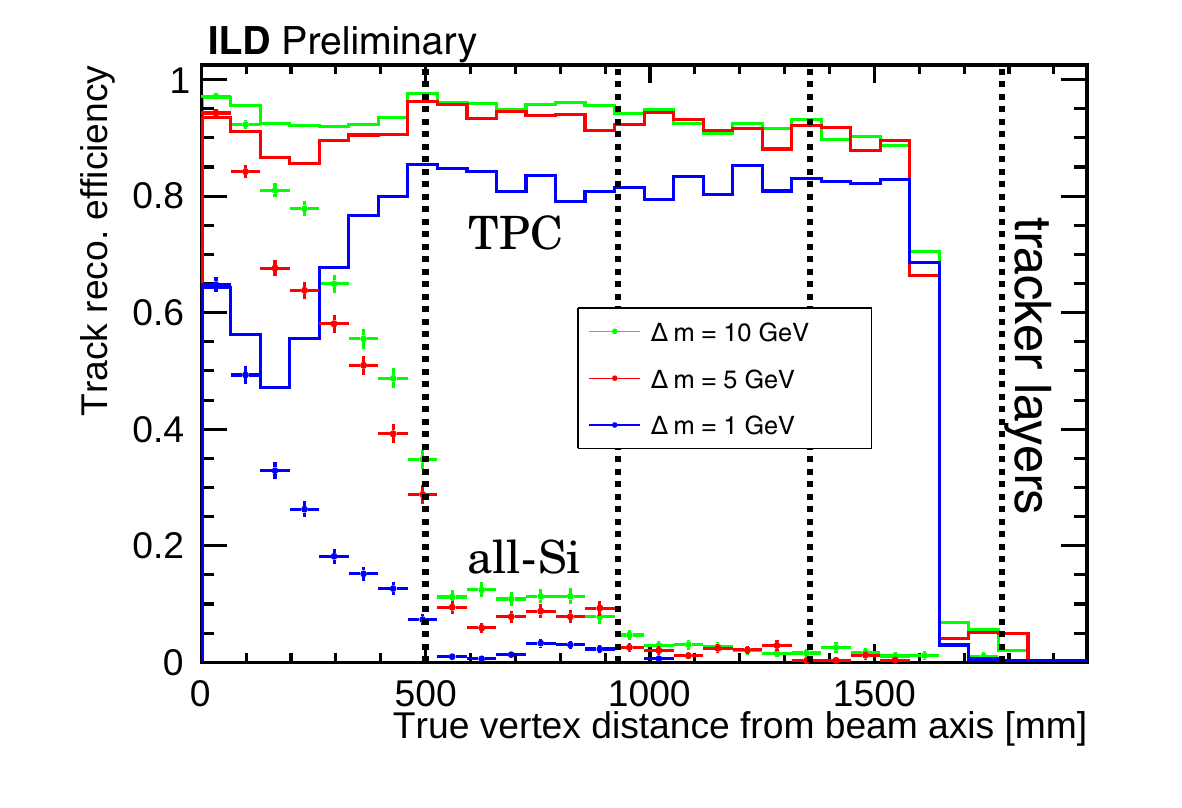}
  \vspace*{\figskip}
  \caption{Track reconstruction efficiency as a function of distance from the beam axis for the heavy scalar samples in the all-silicon ILD design (points), compared to efficiency in the baseline TPC-equipped ILD (solid lines). Different colours correspond to particular scalar mass-splitting scenarios in the test samples. Vertical dashed lines show the position of barrel layers of the outer silicon tracker in all-silicon model. }
  \label{fig:all-si}
\end{figure}

The resulting tracking efficiency (which drives the vertex finding) is presented in Fig.~\ref{fig:all-si} as a function of the decay vertex distance from the beam axis and compared with the corresponding results for the standard ILD design. It can be noted that, if LLP decays inside the vertex detector or inner tracker (the same in both models), the performance is similar. However, for the all-silicon design, the efficiency drops very quickly for larger vertex distances from the beam axis, whereas in the standard ILD design, it stays high almost throughout the entire tracker volume. This is due to the limited number of tracker layers and hence also the number of hits. The latter was required to be at least 4 to reconstruct a track and for this reason the efficiency for the silicon tracker decreases almost to zero already 1\,m from the beam axis.

\section{Conclusions}

Prospects for reconstructing the displaced vertices from LLP decays in the ILD have been studied. An experiment-orientated, model-independent approach has been taken, in which different kinematic properties are analysed, instead of points in the parameter space of a given model. Background from the beam interactions was taken into account, and means of its reduction were identified. For two very challenging and extremely distinctive signatures, the ILD shows good sensitivity for most of the analysed scenarios. Rough estimate of the limit on the signal production cross section is 0.01--0.03\,fb in the considered kinematic region and for decays inside TPC volume. Tests using an alternative, all-silicon ILD design confirm the key role of a TPC in the sensitivity to LLP searches involving soft final states.

\section*{Acknowledgements}
We would like to thank the LCC generator working group and the ILD software working group for providing the simulation and reconstruction tools and producing the Monte Carlo samples used in this study. This work has benefited from computing services provided by the ILC Virtual Organization, supported by the national resource providers of the EGI Federation and the Open Science GRID. The work was supported by the National Science Centre (Poland) under OPUS research project no.~2021/43/B/ST2/01778.


\printbibliography

\end{document}